\documentclass[prb,showpacs,twocolumn,superscriptaddress]{revtex4}

\usepackage[dvipdfm]{graphicx,color}
\usepackage{amsmath, amssymb}
\usepackage{textcomp}
\usepackage{ulem} %for \sout

\begin{document}
\title{Relaxor ferroelectricity induced by electron correlations in a molecular dimer Mott insulator}

\author{S. Iguchi}
\author{S. Sasaki}
\affiliation{Institute for Materials Research, Tohoku University, Sendai, 980-8577, Japan}

\author{N. Yoneyama}
\affiliation{Department of Education, Interdisciplinary Graduate School of Medicine and Engineering, University of Yamanashi, Kohu, Yamanashi 400-8511, Japan}
\affiliation{CREST, Japan Science and Technology Agency, Tokyo 102-0075, Japan}

\author{H. Taniguchi}
\affiliation{Department of Physics, Faculty of Science, Saitama University, Sakura-ku, Saitama, 338-8570, Japan}

\author{T. Nishizaki}
\affiliation{Institute for Materials Research, Tohoku University, Sendai, 980-8577, Japan}

\author{T. Sasaki}
\affiliation{Institute for Materials Research, Tohoku University, Sendai, 980-8577, Japan}
\affiliation{CREST, Japan Science and Technology Agency, Tokyo 102-0075, Japan}

\newcommand{\ka}{$\kappa$-(BEDT-TTF)$_2$Cu$_2$(CN)$_3$}
\newcommand{\be}{$\beta$'-(BEDT-TTF)$_2$ICl$_2$}
\newcommand{\re}[1]{\textcolor{red}{#1}}

\begin{abstract}

We have investigated the dielectric response in an antiferromagnetic dimer-Mott insulator \be{} with square {dimer} lattice, compared to a spin liquid candidate \ka. Temperature dependence of the dielectric constant shows a peak structure obeying Curie-Weiss law with a strong frequency dependence. We found an anisotropic {glassy} ferroelectricity by pyrocurrent measurements, which suggests the charge disproportionation {resulting in an electric dipole} in a dimer. 
{ Each ferroelectric and antiferromagnetic transition temperatures is closely related to the antiferromagnetic interaction energy and a freezing temperature of dipole dynamics in a dimer, respectively.
These correspondences suggest the possible charge-spin coupled degrees of freedom in the system.}

\end{abstract}

\maketitle

%\section{Introduction}

The discovery of new types of multiferroics \cite{1, 2}, where ferroelectricity is driven by nontrivial magnetism, have also initiated new research into a more fundamental understanding of spin-charge coupled phenomena in strongly correlated systems. However, controlling ferroelectricity is difficult in such systems due to strong spin structure-electrical dipole coupling. Thus, electronic ferroelectricity \cite{3, 4} is now explored as an alternative route towards multiferroics, related to the charge ordered state in Mott insulators \cite{3, 5}. 
Among them, molecular dimer Mott insulators are a good candidate for the ferroelectricity, which are typical strongly correlated electron systems, constructed by dimers of organic molecules such as bis(ethylenedithio)tetrathiafulvalene (BEDT-TTF). When each dimer in a crystal possesses a charge carrier, in particular for the half-filled case in a dimer molecular orbital, the electron system often shows a Mott transition for relatively large (on-site) Coulomb interactions, $U$, at the dimer site compared to the electron hopping energy, $t$ \cite{6}. An important physical insight is that a dimer, rather than a molecule, can be viewed as the primitive constituent in a crystal. In this way, the understanding of strongly correlated electron systems based on the dimer picture within the Hubbard model has progressed \cite{7, 8}. However, recent reports on dielectric properties \cite{9} imply that the charge degrees of freedom in a dimer survive even in the dimer Mott insulating phase \cite{10, 11}. This discovery will enable a deeper understanding of organic Mott insulators beyond the successful dimer lattice model and the discovery of new functionality. The origin of this ferroelectricity is different from the recently discovered hydrogen-bonded organic ferroelectrics \cite{12} and the Peierls instability-induced inversion symmetry breaking in one-dimensional systems \cite{13}.

An anomaly in the dielectric constant has been observed for a dimer Mott insulator \ka{}  \cite{9} with a {\it triangular} lattice in terms of the dimer picture, which is one candidate for a spin liquid \cite{14, 15, 16}. The temperature dependence of the dielectric constant for \ka{}  shows a ferroelectric relaxor-like frequency dependence, expected to exhibit charge disproportionation \cite{11} in a dimer or strong charge fluctuations near a charge ordered state \cite{10}. The dielectric constant in \ka{}  obeys the Curie Weiss law with a Curie temperature of 6 K. However, there is no explicit evidence of ferroelectricity or the charge freezing typically seen alongside relaxor behavior. This dielectric anomaly may be related to the quantum spin liquid nature of this material, which further complicates the physical origin of the dielectric constant anomaly. Here, we have observed a similar dielectric anomaly due to the response of charge degrees of freedom in a dimer for a simpler dimer Mott insulator \be{} with a square lattice and discovered evidence for relaxor ferroelectricity by pyrocurrent measurements with an anisotropy consistent with the dimer arrangement. These results may lead to the future discovery of electronic ferroelectric materials in dimer Mott insulators and a deeper understanding of the spin liquid state.

 \be{} is an anisotropic quasi-two dimensional molecular dimer Mott insulator with one hole carrier in the dimer of a BEDT-TTF molecule. The molecular dimer forms a square lattice, in contrast to the triangular lattice in \ka. 
The space group of this system is triclinic P$\bar{1}$ where the inversion symmetry produces one dimer in a unit cell, as shown in Fig. 1(a) \cite{17}. The dimers are aligned along the $b$ axis, arrayed in a column on the $bc$ plane, and the layers are stacked along the $a^*$ axis ($\perp bc$ plane), well separated by ICl$_{2}$ anion layers.
Dimerization was supported by NMR \cite{18} and a calculation of the intradimer transfer integral \cite{17}. Thus, the dimer picture is applicable for this system, and this material is considered {to be} a half-filled Mott insulator with an antiferromagnetic transition at $T_{\rm N} = 22$ K {and antiferromagnetic interaction $J_{\rm AF}=59$ K}\cite{19}. The  long-range antiferromagnetic order below $T_{\bf N}$ indicates that this system is located inside the well-defined dimer Mott insulating region with large $U/t$ value \cite{6}. 
Although band structure calculations \cite{17} and optical conductivity measurements \cite{20,21} indicated strong one-dimensional anisotropy along the $b$ axis (in the $bc$ plane), anisotropic two-dimensional feature was reported in transport \cite{Tajima2008} and magnetism\cite{19}. The system becomes superconducting under 8 GPa pressure \cite{17}, as typically seen in dimer Mott systems by reducing $U/t$ through the Mott insulator-metal/superconductor transition \cite{8}. Thus, this system could be understood as a typical quasi-two-dimensional dimer-Mott insulator 
different from the charge order system with ferroelectricity as seen in quasi-one dimensional (TMTTF)$_{2}X$ system \cite{13}.

%%%%%%%%%%%%%%%%%%%%%%%%%%
% fig 1
\begin{figure}
\includegraphics*[width=8.3cm]{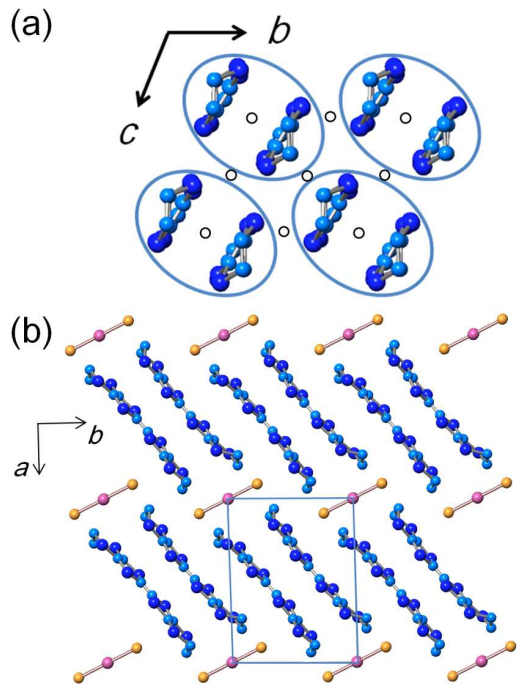}
\caption{(Color online) Crystal structure of \be. (a) BEDT-TTF molecule arrangement viewed along the long axis of BEDT-TTF molecules and (b) the crystal structure seen from the $c$ axis. The blue ovals represent dimerized molecules symmetric with respect to the inversion operation at the point represented by open circles. The unit cell is represented by blue lines in (b). Hydrogen atoms are not drawn for clarity.}
\label{fig1}
\end{figure}
%%%%%%%%%%%%%%%%%%%%%%%%%

%\section{Experimental}

The samples used in this study were grown by an electrochemical method \cite{22}. Typical dimensions of the specimens are 0.7 mm $\times$ 0.7 mm $\times$ 2 mm along the $a^*$, $b^*$, and $c$ axes, respectively. An electric field of 5 V/cm was applied along different crystallographic directions in order to measure the dielectric constant with an LCR meter. Pyrocurrent measurements were performed with increasing temperature at a rate of 10 K/min after applying a poling electric field from -1.2 kV/cm to 4.8 kV/cm at approximately 100 K. 

%\section{Results}

Figure 2 shows the temperature dependence of the dielectric constant $\varepsilon$  for \be{} in (a) $E \parallel a^*$ perpendicular to the two dimensional BEDT-TTF layer and (b) $E \parallel b$ parallel to the dimer array at each frequency (20 to 200 kHz). An anomaly with a strong frequency dependence is discerned in both directions at around 80-150 K. The anomalies at high frequencies, e.g.  $\geq 50$ kHz, show up at 130-150 K. With decreasing frequency, the anomaly in the dielectric constant becomes large and forms a peak structure. A similar frequency dependent dielectric constant has been observed for \ka, which is typically seen in relaxor ferroelectric materials with randomness and frustration. In \be{} at 500 Hz, the peak value from the background level at the lowest temperature corresponds to a $\Delta \varepsilon=10$ which is also comparable to that in \ka{} \cite{9} at the corresponding frequency and crystallographic direction. While we could not obtain reliable data for 
$\varepsilon$ at frequencies lower than 500 Hz in $E \parallel a^*$, we were able to measure $E \parallel b$. At the lowest frequency of 20 Hz, the anomaly grows as large as $\Delta \varepsilon = 20$ for $E \parallel b$. Thus, dielectric anomalies discerned in the both directions are of an identical origin, which is expected as charge disproportionation occurs in a dimer along the direction from the center of a BEDT-TTF molecule to the other one in the dimer.

%%%%%%%%%%%%%%%%%%%%%%%%%%
% fig 2

\begin{figure}
\includegraphics*[width=8.2cm]{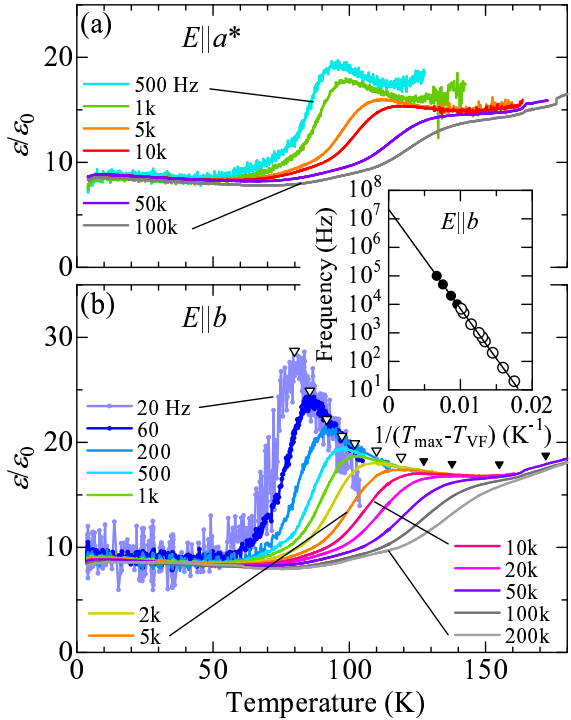}
\caption{(Color online) Temperature dependence of the dielectric constant, $\varepsilon$, in (a) $E \parallel a^*$ ($\perp bc$ plane) at frequencies from 500 Hz to 100 kHz and (b) $E \parallel b$ from 20 Hz to 200 kHz for \be. The data in the high temperature region was omitted for clarity. The inset shows the Vogel-Fulcher (VF) relation using the dielectric constant at frequencies from 20 Hz to 7 kHz indicated as open triangles in (b). The black solid line in the inset is a best fit to the VF relation. $T_{\rm max}$ and $T_{\rm VF}  = 23$ K represent the temperatures at which $\varepsilon$ takes a maximum value at each frequency and the charge freezing temperature obtained by the VF fitting, respectively. Solid triangles in (b) represent the expected $T_{\rm max}$ by the relation indicated the filled circles in the inset. }
\label{fig2}
\end{figure}
%%%%%%%%%%%%%%%%%%%%%%%%%%

The inverse dielectric constant subtracted by a constant is shown in Fig. 3(a), which clearly obeys the Curie-Weiss law with a constant term $\varepsilon_{\rm const.}$:
$
\varepsilon/\varepsilon_{\rm 0} = \frac{C}{T-T_{\rm C}} + \varepsilon_{\rm const.}/\varepsilon_{\rm 0}.
$
Here $\varepsilon_{0}$, $C$, and $T_{\rm C}$ are the vacuum permittivity, Curie constant, and Curie temperature, respectively. The black line in Fig. 3(a) represents the best fit obtained for the data at frequencies less than 5 kHz. The constant term $\varepsilon_{\rm const.}/\varepsilon_{0}$ is 13.8. According to the slope of the inverse dielectric constant, $C$, the dipole moment parallel to the $b$ axis is estimated to be 0.13$ed$, where $e$ and $d$ are the electron charge and distance between the BEDT-TTF molecules in a dimer, 3.6 \AA \cite{17}, respectively. The Curie temperature $T_{\rm C}$ was found to be 67 K. This $T_{\rm C}$ is ten times larger than that for \ka{} {in which large spin and charge frustrations have been expected \cite{9, 14, 15, 16}.}

While the apparent ferroelectric Curie temperature $T_{\rm C}$  and relaxor-like charge-freezing or Vogel-Fulcher temperature, $T_{\rm VF}$, of the domains reported for \ka{} are identical at 6 K \cite{9}, these temperatures for \be, as described below, are clearly different as shown in Fig. 3(a) and the inset of Fig. 2, respectively. The inset of Fig. 2 shows a fit to the data for $E \parallel b$ of \be{} to the Vogel-Fulcher (VF) relation for $\varepsilon$:
$
f = f_{0}\exp{\bigl\{ -E_{\rm VF}/(T_{\rm max}-T_{\rm VF})\bigr\}}
$
where $f_{0}$, $E_{\rm VF}$, and $T_{\rm max}$, are the representative flipping frequency of electric dipoles, average random energy barrier, and temperature at which $\varepsilon$ takes a maximum, respectively. While fitting to this relation, $\varepsilon$ as measured at frequencies from 20 Hz to 7 kHz in Fig. 2(b) was used. 
The empirical VF relation well represents the frequency dependence of the peak in $\varepsilon$  with a $T_{\rm VF} = 23$ K, $E_{\rm VF} = 804$ K, and  $f_{\rm 0}=2.17 \times 10^{7}$ Hz. In this system, the charge freezing temperature, $T_{\rm VF}$, is much different from the Curie temperature $T_{\rm C}$ = 67 K, although these temperatures have similar meaning and are often considered to be the same. Since the VF relation reveals little about the temperature dependence of the dielectric constant, we measured the pyrocurrent to uncover the actual ferroelectric transition temperature.

%%%%%%%%%%%%%%%%%%%%%%%%%%
% fig 3
\begin{figure}
\includegraphics*[width=8.2cm]{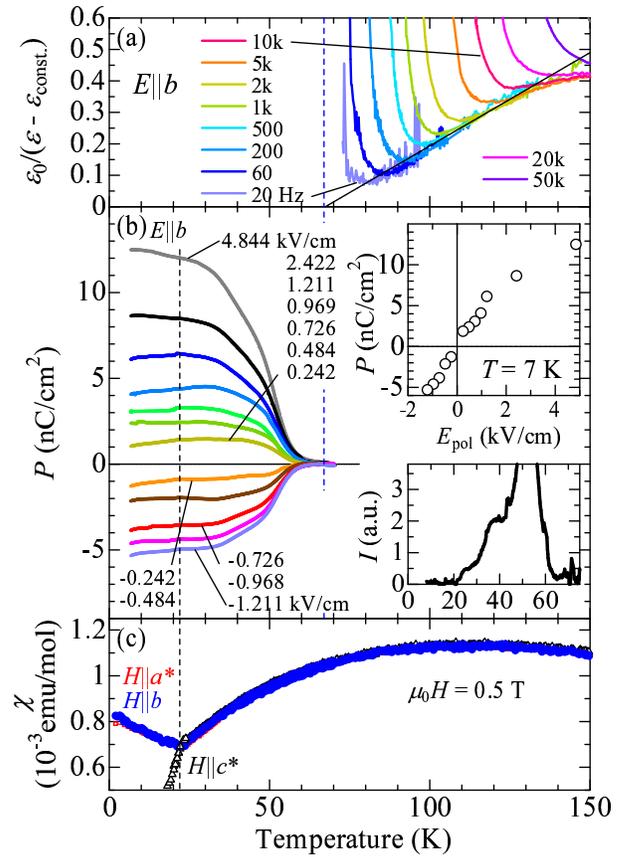}
\caption{(Color online) (a) Temperature dependence of inverse dielectric constant at each frequency in \be
, {where $\varepsilon_{\rm const.}/\varepsilon_{0} = 13.8$} The black straight line represents the best fit result to the Curie-Weiss law. Temperature dependence of (b) polarization with $E \parallel b$ and a poling electric field from -1.211 to +4.844 kV/cm obtained by pyrocurrent measurements, and (c) magnetic susceptibility with $H \parallel a^*$, $b$ and $c$ for \be. The polarization at 7 K for each poling field in (b) is plotted in the {upper} inset, which does not correspond to the standard $P-E$ curve at 7 K. A typical pyrocurrent for $E_{\rm pol}=2.422$ kV/cm is shown in the lower inset. }
\label{fig3}
\end{figure}
%%%%%%%%%%%%%%%%%%%%%%%%%%

Figure 3(b) shows the electric polarization, $P$, with $E \parallel b$ obtained by pyrocurrent measurements and (c) the temperature dependence of the magnetic susceptibility, $\chi$, in a magnetic field with $H \parallel a^*$, $b$, and $c$ for \be. A small but measurable polarization was observed {below} a critical temperature of $T_{\rm FE}$ = 62 K, which is nearly identical to the estimated $T_{\rm C}$ = 67 K. 
{As shown in the lower inset of Fig. 3(b) the pyrocurrent starts to appear at $T_{\rm N}$, increases gradually, takes a maximum just below $T_{\rm FE}$ and disappears at $T_{\rm FE}$ in warming process. $P$ starts to decrease at $T_{\bf N}$ with appearing the pyrocurrent.}
$P$ {at the lowest temperature increases linearly with} poling electric field,  $E_{\rm pol}$, at least up to 2 kV/cm. Leakage currents prevented the measurement of $P$ for $|E_{\rm pol}|$ larger than 5 kV/cm. { A pyrocurrent while poling with $E \parallel a^*$ could not be observed up to 3 kV/cm, which supports the assertion that electric dipoles align along the dimer arrangement.} {The expected full moment is estimated as 2 $\mu$C/cm$^2$  assuming the dipole moment of 0.13$ed$, which is comparable to hydrogen bonded ferroelectric materials \cite{12}. However, the experimentally obtained polarization of 13 nC/cm$^2$ at $E_{\rm pol}$ = 4.8 kV/cm is small.}
The electric field linearity and small polarization values indicate that the fully polarized state has not yet been obtained under poling electric fields of 5 kV/cm. Thus, such a polarization indicates the formation of glassy polar domains, also consistent with the strong frequency dependence in $\varepsilon$  (Figs. 2(a) and 2(b)). The small and glassy polarization by the charge disproportionation 
in this compound might be overlooked in NMR\cite{18, 25} and the other measurements. 
The temperature dependence of $\chi$ (see Fig. 3(c)) does not exhibit a notable anomaly at $T_{\rm FE}$, perhaps because the glassy ferroelectricity observed here is only electric field-induced. 
The magnetic phase transition at $T_{\rm N}$ is thermodynamically well defined and may affect the electric dipoles different from the glassy transition. 
{The fact that the pyrocurrents do not flow below $T_{\rm N}$ indicates that the dynamics of charge disproportionation is strongly suppressed by the long-range antiferromagnetic order and the charge glassy state relaxes above $T_{\rm N}$.}
Therefore, in this system, the spin and charge degrees of freedom within a dimer have a moderate coupling, even though the antiferromagnetic and ferroelectric transition temperatures are different. 

%\section{Discussion}

Our discovery of ferroelectricity with strong frequency dependence, similar to that in \ka, may be due to the intrinsic phenomenon of electronic ferroelectricity in a dimer-Mott insulator with antiferromagnetic interactions. Although the mechanism for slow dynamics with possible domain formation is an open issue, the discrepancy between the actual ferroelectric $T_{\rm FE}$ ($\simeq T_{\rm C}$) and the fit result of $T_{\rm VF}$ {can be attributed to} 
spin-charge coupling as shown below. The charge distribution in the dimer-Mott insulating state has the center of inversion as shown in Fig. 4(a). However, ferroelectricity or charge disproportionation in a dimer is driven intrinsically by the hybridization of two orbitals with different parity \cite{23}. Such partial hybridization produces electrically polar excited states, considered to result from collective phenomena induced by the inter dimer Coulomb correlation, $V$ (Fig. 4(b) and 4(c)). This interaction can lead to charge fluctuations in a dimer, as predicted theoretically \cite{10, 11}, showing a Curie-Weiss behavior for the dielectric constant. However, when the temperature of the system decreases below the energy scale of the antiferromagnetic interaction $J_{\rm AF}$ (= 59 K \cite{19}) {at which temperature $1/(T_{1}T)$ in $^{13}$C NMR measurement starts to increase \cite{Tajima2008}}, the approach of electrons with opposite spins is favorable due to the increase in transfer as  $J_{\rm AF}= -t^2/U$ (Fig. 4(d)), which may be relevant to the increase in $T_{\rm N}$ \cite{24} and the presence of superconductivity under pressure \cite{17}. Thus, an antiferroelectric interaction is possibly realized with $J_{\rm AF}$ rather than a ferroelectric interaction without $J_{\rm AF}$. In this sense, $T_{\rm VF}$ is almost the same as $T_{\rm N}$ and is related to the antiferroelectric interaction between ferroelectric domains. Therefore, additional randomness and frustration between ferroelectric and antiferroelectric interactions are produced by the spin-charge coupled degrees of freedom, resulting in a finite polarization at $T_{\rm FE} \simeq T_{\rm C}$ higher than $T_{\rm VF}$.

%%%%%%%%%%%%%%%%%%%%%%%%%%
% fig 4
\begin{figure}
\includegraphics*[width=8.0cm]{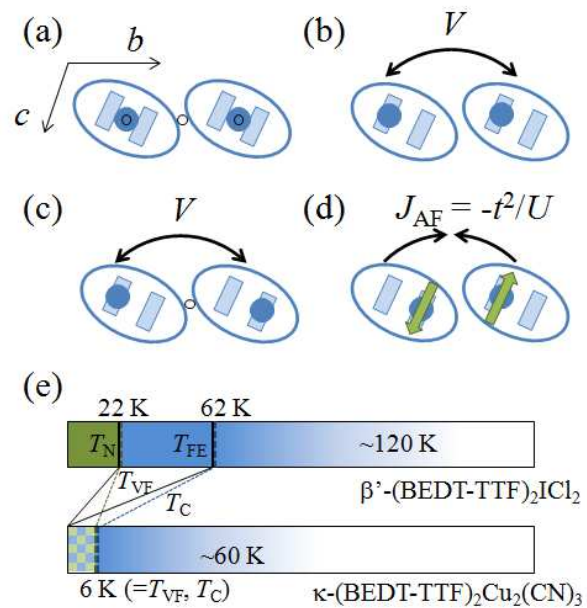}
\caption{(Color online) Schematic pictures of (a) dimer-Mott insulator, inter dimer Coulomb correlation ($V$) with (b) ferroelectric and (c) antiferroelectric dimers, and (d) antiferromagnetic interaction ($J_{\rm AF}$) with antiferroelectric dimers within the $bc$ layer, where a dimer is represented by an open circle. Solid circles and arrows represent charge and spin, respectively. (e) Schematic electronic phase diagram for \be{} and \ka, where the dielectric constant starts to increase at around 120 K and 60 K, respectively. $T_{\rm N}$ and $T_{\rm FE}$ correspond to antiferromagnetic and ferroelectric transition temperatures, respectively. The change in parameters of the Curie-Weiss ($T_{\rm C}$) and the Vogel-Fulcher ($T_{\rm VF}$) temperatures are also shown.}
\label{fig4}
\end{figure}
%%%%%%%%%%%%%%%%%%%%%%%%%%

Figure 4(e) shows a schematic picture of the electronic phases with respect to the parameters $T_{\rm VF}$ and $T_{\rm C}$ in $\beta$'-(BEDT-TTF)$_2$ICl$_2$ and $\kappa$-(BEDT-TTF)$_2$Cu$_2$(CN)$_3$. The discrepancy in the phases and parameters is clearly seen between the crystals. In the case of $\kappa$-(BEDT-TTF)$_2$Cu$_2$(CN)$_3$, $T_{\rm C}$ and $T_{\rm VF}$ are the same, implying that the strong competition between ferroelectric and antiferroelectric interactions, where magnetic and electric phase transitions have not been observed at any temperature. 
In the dipole-dimer picture, $\kappa$-type structure has two differently directing dimers in a BEDT-TTF layer.
Thus, two different directions of polarization are possible \cite{10, 11} and can compete with each other even by ferroelectric interaction. Moreover, antiferroelectric interaction due to antiferromagnetic one{, if any,} also competes or fluctuates due to the spin frustration in this spin-liquid candidate. 
{It is interesting to compare with a report on a ferroelectricity appeared with an antiferromagnetic long-range order at $T_{\rm N}$ in $\kappa$-(BEDT-TTF)$_{2}$Cu[N(CN)$_{2}$]Cl{\cite{25}.}}
{Different explanation for the dielectric anomaly in $\kappa$-type compounds was reported quite recently \cite{27}, where the tilt of BEDT-TTF molecules and their coupling to the anion layers were supposed for a probable origin. Further experimental investigations are necessary to understand the dielectric anomalies considering the charge-spin composite degrees of freedom in the $\pi$-orbitals of the molecular dimer-Mott system.}

%\section{Summary}

In conclusion, we have revealed the dielectric response in the antiferromagnetic dimer-Mott insulator \be{} with a square lattice, compared to the spin liquid candidate \ka. We observed a peak structure in $\varepsilon$ with strong frequency dependence, probably due to ferroelectric relaxor-like domain formation. The anisotropic pyrocurrent corresponds closely with charge disproportionation in a dimer, and shows the ferroelectric transition temperature is {possibly} related to the antiferromagnetic interaction. This, in turn, supports {a possible scenario of} the finite spin-charge coupled degrees of freedom in the system. Our result 
{implies} the importance of the charge degrees of freedom
 in BEDT-TTF dimers, as a key concept widely applicable to molecular dimer Mott transition systems. 

%\section*{Acknowledgment}

The authors would like to thank S. Ishihara, S. Iwai, and I. Terasaki for valuable discussions, and acknowledge the experimental contributions by M. Abdel-Jawad at the initial stage of this study. This work was supported by Grants-in-Aid for Scientific Research from JSPS (24540357) and MEXT, Japan (23110702 and 23102502) and a research grant from The Murata Science Foundation.

%References

\end{document}